\documentclass[prl,superscriptaddress,preprint]{revtex4-2}
\usepackage{graphicx}
\usepackage{bm}
\usepackage{amsmath}
\usepackage{color}
\usepackage[latin9]{inputenc}
\usepackage{amstext}
\usepackage[normalem]{ulem}
\usepackage{soul}
\usepackage{natbib}
\usepackage{setspace}

\begin{document}

\title{On the origin of \\ the unusual strain morphologies and polar Moir\'e patterns \\ in twisted ferroelectrics}

\author{Sergey Prosandeev}

\affiliation{Smart Ferroic Materials Center, Department of Physics and Institute for Nanoscience and Engineering, University of Arkansas, Fayetteville, Arkansas 72701, USA}

\author{Charles Paillard}

\affiliation{Smart Ferroic Materials Center, Department of Physics and Institute for Nanoscience and Engineering, University of Arkansas, Fayetteville, Arkansas 72701, USA}

\affiliation{Universit\'{e} Paris-Saclay, CentraleSup\'{e}lec, CNRS, Laboratoire SPMS, 91190, Gif-sur-Yvette, France}

\author{L. Bellaiche}

\affiliation{Smart Ferroic Materials Center, Department of Physics and Institute for Nanoscience and Engineering, University of Arkansas, Fayetteville, Arkansas 72701, USA}

\affiliation{Department of Materials Science and Engineering, Tel Aviv University, Ramat Aviv, Tel Aviv 6997801, Israel}

\begin{abstract}
Density functional theory calculations are conducted to understand and reveal the origin of the complex shear strain morphology and of the polar Moir\'e  topological pattern recently observed in  twisted BaTiO$_3$ bilayers.
Our first-principles calculations, along with an original analysis of them allowing the decomposition
of forces into the acoustic and optical contributions, point out to the occurrence of forces mostly acting on the {\it acoustic-related} motions to produce the standing  waves of the shear strain.  Such acoustic waves naturally generate a striking self-organization of the shear strains, and hence create a peculiar gradient of these shear strains. A Moir\'e dipole pattern, consisting of the interpenetrated arrays of vortices and antivortices made of the electric dipoles, then  mostly arises due to the coupling of this gradient of the shear strain with the electric dipoles. Furthermore, other forces, namely acting on the motions associated with  the {\it optical phonons}, could also play a role in the formation of these polar vortices and antivortices, but at a smaller extent.
 \end{abstract}

\maketitle

\newpage

Twistronics is a new  quickly developing field of study of twisted atomic layers of materials to discover novel and peculiar properties  \cite{Carr}. Studies in that field began with the investigation of the interference of light on some twisted bilayers \cite{photonic}. The resulting interference structure was called Moir\'e pattern and can contain intricate features, including chirality, being dependent on the value of the twisting angle.
Twistronics was also investigated in 2D Van der Waals materials such as graphene, hexagonal boron nitrides, or transition metal dichalcogenides~\cite{He2021}, with inherent Moir\'e pattern made of the interpenetrated periodic arrays of the metallic and dielectric regions. Many unusual electronic properties, that are promising for specific technologies, were found there. Examples include tunable the correlated states~\cite{tunable} and Chern insulator phases~\cite{Jia,Herzog} in graphene, and even superconductivity at some magic angle \cite{Andrei}. Moir\'e patterns have also been recently advocated to be relevant to neuromorphic applications \cite{Yan,Fan}, with the possibility of gate-controlled neuromorphic transistors promising reconfigurable synaptic response and adaptive learning at the room temperature.

Very recently, such twisted atomic layers were also engineered in dielectric ferroelectric materials; especially  in bilayers of BaTiO$_3$ (BTO) \cite{jorge}, where two kinds of striking morphologies were observed:
 (1) Moir\'e patterns consisting of the vortices and antivortices made of  electric dipoles, along with (2) an unusual self-organization of the inhomogeneous shear strains consisting of four different types of quadrants. Interestingly, when reading the literature, it is not clear what is the main feature and what is the secondary driven effect among items (1) and (2) or if these two items have equal footing, when twisting ferroelectric layers. One may even wonder if there are other and even more important mechanisms that are responsible for  the simultaneous occurrence of items (1) and (2). As a matter of fact, different conclusions can be drawn from previous  investigations. For instance, some authors of Ref. \cite{jorge} performed first-principle calculations imposing vortices and antivortices of electric dipoles as the starting point and then allowed the strain to relax within this imposed polar Moir\'e pattern. The resulting strain was found to be made precisely of four quadrants, exactly as in the measurements of the same article \cite{jorge}, which tends to suggest that the twisted oxide layers first create this specific polar pattern, which then generates the unusual shear strain morphology via dipole-strain couplings (such as flexoelectricity ~\cite{Kogan1964,Zubko2013}).  In contrast, Reference \cite{flexo} assumed that the twisted BTO layers first induce the aforementioned specific morphology of the shear strain (and of its associated gradients), which was then numerically proven to induce the formation of the polar vortices and antivortices via flexoelectricity. Moreover,  Lee {\it et al.} adopted another thinking by assuming that there is a magic angle at which the electronic bands flatten, and that the Moir\'e polar pattern arises because of the intimate connection between these electronic bands, via the localization of the electrons, and the electric dipoles \cite{Souza}.

 There are therefore contradictory views on the origin of the unusual strain and dipolar patterns in  twisted dielectric materials, while such materials were predicted  \cite{flexo} to exhibit striking features such as the control of the size and sense of the rotation of the polar vortices, space-dependent gyrotropic properties and Beresinskiy-Kosterlitz-Thouless-like phase transitions \cite{Kosterlitz-Thouless,Berezinskii}. Determining such origin is therefore highly desired to deepen fundamental knowledge on topology, twistronics, and ferroelectrics but also for possible novel applications.
Additional investigations are thus currently needed.

In the present study, we performed density functional (DFT) calculations, and conducted original analysis, to answer the question about the origin of the strain and polar patterns in  twisted ferroelectrics. We  numerically find that the  twisted layers in BTO yield the emergence of   forces mostly acting on the displacements associated with the shear strain. Considering such forces as the source of the acoustic atomic displacements naturally  results in the occurrence of the shear strains possessing the four aforementioned quadrants. The formation of the polar vortices and antipolar vortices then mostly arises from the flexoelectric coupling between this shear strain and electric dipoles. Forces acting on the motions associated with the optical phonons can play an additional but smaller role in the occurrence of the  polar topological defects.

Here, we  conduct  density-functional-theory (DFT) calculations \cite{Kohn}  for the twisted BaTiO$_3$ (BTO) bilayers. The Vienna Ab Initio Software package (VASP)~\cite{Kresse1994,Kresse1996} with the Projector Augmented Waves (PAW) pseudopotentials~\cite{Blochl1994,Kresse1999} and  PBE exchange-correlation~\cite{Perdew1996} is employed. We use  a BaTiO$_3$ supercell that is initially (before twisting) periodic along any Cartesian direction and  has a $4\times 4\times 2$ dimension, in terms of the 5-atom  primitive unit cell lattice constant equal to 3.985 \AA. This latter minimizes the internal energy of BTO bulk within the cubic paraelectric crystal structure. It will be denoted as $a_{lat}$ in the following. Notice that the $4\times 4\times 2$ supercell can be considered to possess two (001) BTO bilayers. This is why, below, we distinguish between these two layers by denoting them as the bottom atomic layer and the top atomic layer. We set a plane wave cut-off of 550~eV, and use a  $2\times2\times4$ sampling of the first Brillouin zone.  The electronic self-consistent cycle was considered converged when differences in energy were lower than $10^{-8}$~eV.

\begin{figure}[htbp]
{\includegraphics[scale=0.5]{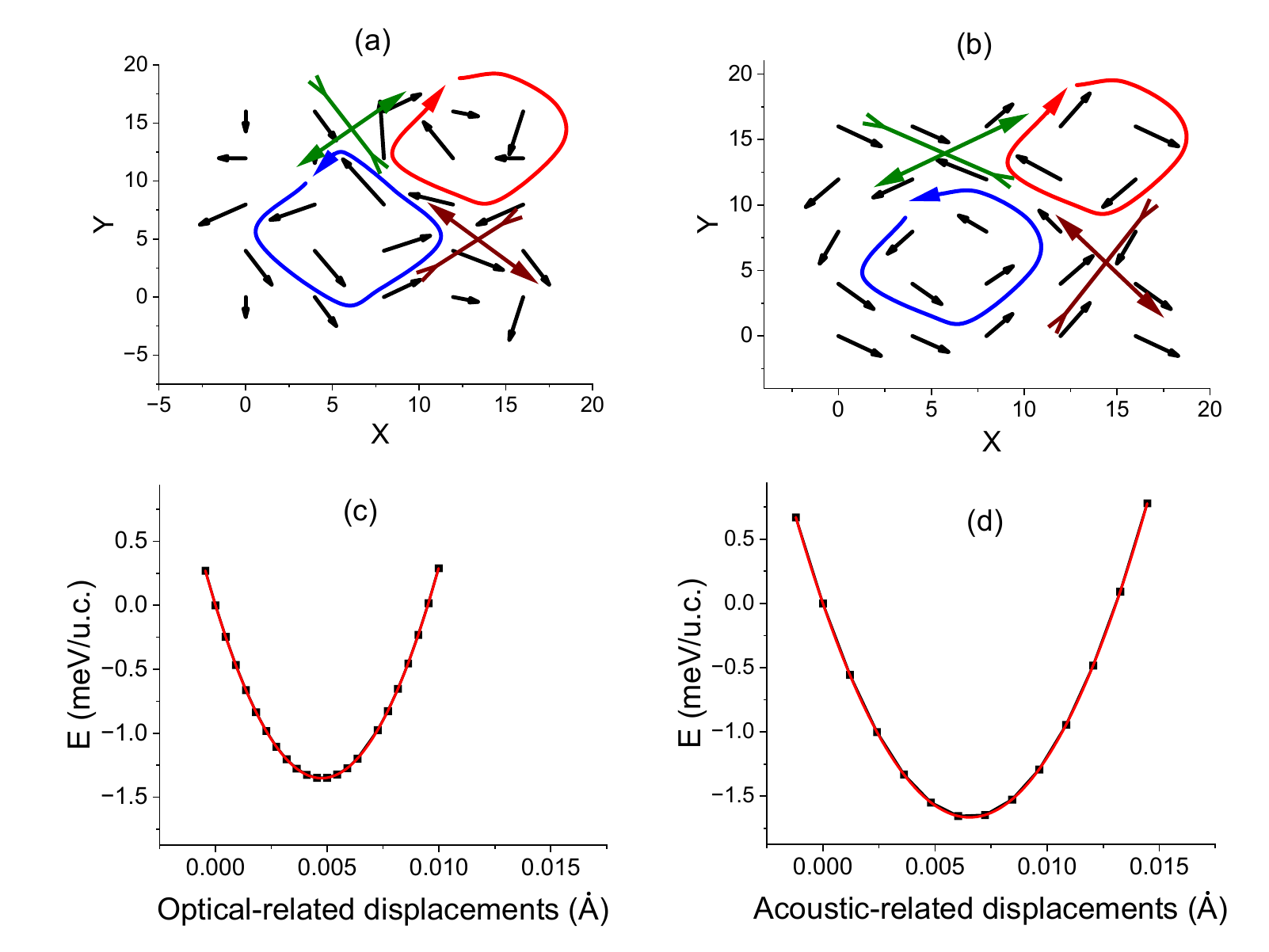}}
\caption{(Color online) Forces centered on the Ti ions in the bottom atomic layer of the BTO twisted bilayer in the  4$\times$4$\times$2 supercell, and related energies. (a) The ``optical'' forces and (b) The ``acoustic''  forces for the angle of twisting of 1$^\circ$.  (c) The energy per 5-atom when varying the displacement of the optical mode. (d) The energy per 5-atom  when varying the displacement of the acoustic mode. In panels (a) and (b), the locations of the two types of vortices are marked in red and blue while those of the antivortices are marked in green and brown. The amplitude of the acoustic force of Panel b is about 3.70 times larger than that of the optical one of Panel a.}
\label{Fig1}
\end{figure}

\begin{figure}[htbp]
{\includegraphics[scale=0.7]{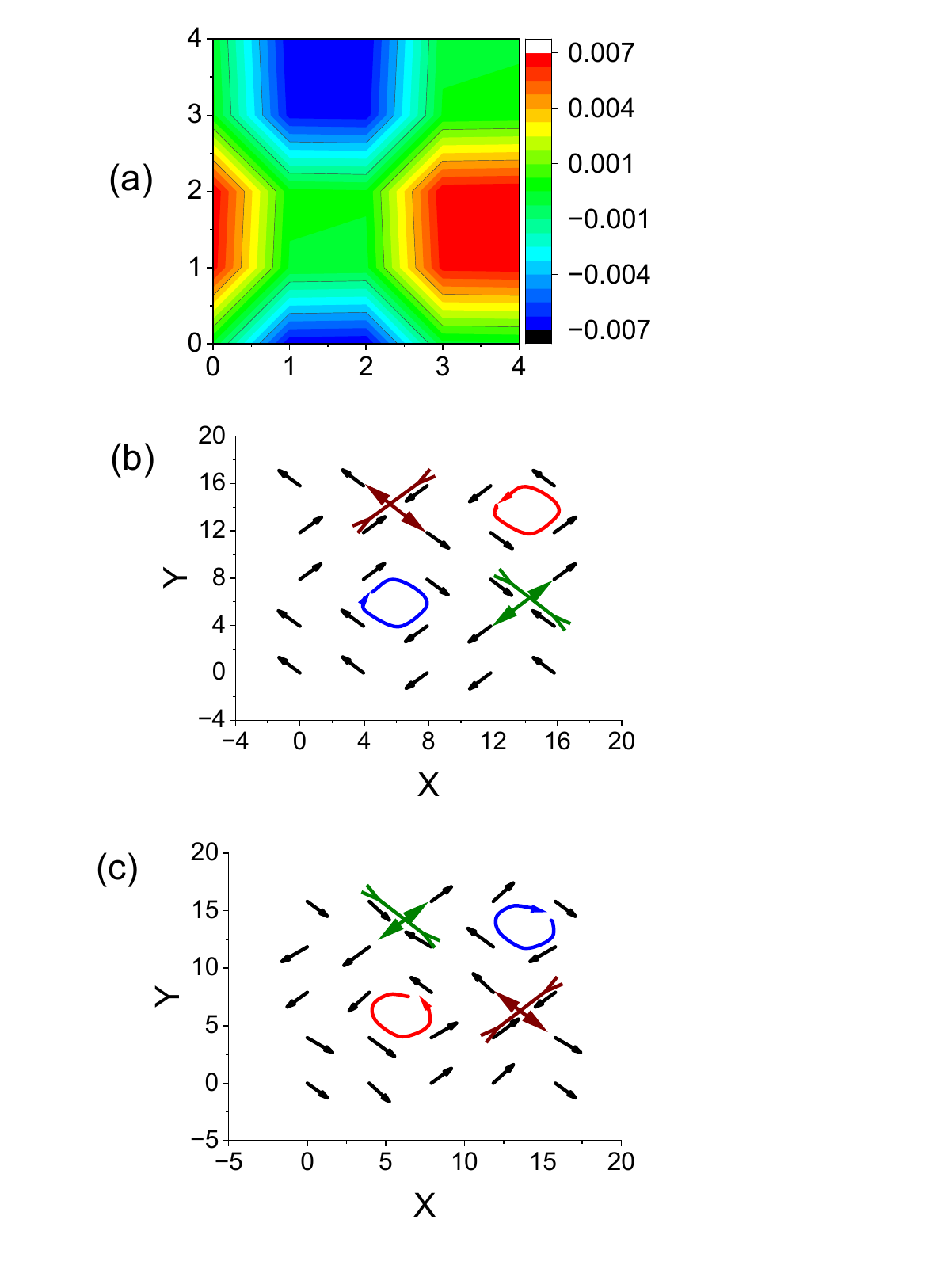}}
\caption{(Color online) Map in a (x,y) plane of the shear strain morphology from Eq. (7) (panel a), of the polarization pattern  from Eq. (\ref{polarizationfinal}) (panel b) and of the polarization self-organization  from Eq. (\ref{polarizationfinaloptical}) (panel c)
with  selected values of $L$, $\mu_{xyxy}$, $k_{acoustic}$, $k_{optic}$,  $Z^*$, $a_{lat}$ as well as for the $A$ and $\phi$ coefficients (see text).  The locations of the two types of vortices are marked in red and blue while those of the antivortices are marked in green and brown in panels b and c. }
\label{Fig2}
\end{figure}

We artificially rotate the positions of the Ti ions in the top atomic layer by  an angle $\theta$ in the supercells, such as the new  Cartesian components of the  coordinates are:

\begin{eqnarray}
x'_i=cos(\theta) x_i-sin(\theta)y_i \nonumber \\
y'_i=sin(\theta) x_i+cos(\theta)y_i
\end{eqnarray}
where $(x_i,y_i)$ are the initial (x,y) coordinates in the top plane, that are also equal to the initial and final coordinates in the bottom plane, while $(x'_i,y'_i)$ are the final coordinates in the twisted (top) plane.  The origin of the coordinates in this study  is located at the center of the square made of four neighboring Ti ions within the top BTO layer. After this twist is done, the forces acting on the ions become finite, with these forces being calculated from the Hellmann-Feynman theorem \cite{Hell}. These forces show the directions of the desired distortions due to twisting.

To analyze these forces, we separate them into optical and acoustic contributions following a strategy often used within the effective Hamiltonian schemes \cite{Vanderbilt-PRL,Vanderbilt-PRB,Bellaiche}.  For instance, let us select one specific Ti atom, and call its force along $x$ as F$_{Ti,x}$. We now look at the two nearest oxygen ions being located on the right and left sides of this Ti ion along the x-axis,  respectively, and sum the $x$-component of their force.  We then divide such sum by two, and call it $F_{Opara,x}$. We also look at the four closest oxygen ions that are in the same $(y,z)$ plane than this Ti atom,  sum their forces along $x$, and divide the result by four. The resulting value is coined as F$_{Operp,x}$. We also look at the eight Ba ions that are closest to this Ti ion, and sum their force along $x$. We then divide this sum by eight, and call it $F_{Ba,x}$. One can then determine a force along $x$ that is acting on the so-called local mode (hence on the electric dipoles and which is related to optical phonons) via:

\begin{eqnarray}
F_{optical,x}= \chi_{Ba} F_{Ba,x} + \chi_{Ti} F_{Ti,x}+\chi_{Opara}  F_{Opara,x} + 2 \chi_{Operp} F_{Operp,x}
\label{Foptical}
\end{eqnarray}
with the $\chi$ coefficients corresponding to the soft-mode eigenvector of the force-constant matrix of BTO. These coefficients have been given in Table VII of Ref. \cite{King-Smith}, and are: $\chi_{Ba}=0.20$, $\chi_{Ti}=0.76$, $\chi_{Opara}=-0.53$ and $\chi_{Operp}=-0.21$. A similar calculation can be done for the forces along $y$ and $z$ for that Ti ion, and therefore one can get the vectorial force felt by the local mode centered on this Ti ion.

Then, we can also determine the $x$-component of the force centered on a specific Ti ion and that is now related to the acoustic mode (that is related to the strain). For that, one has to now use the following formula:

\begin{eqnarray}
F_{acoustic,x}= \frac{1}{\sqrt{5}} \left [ F_{Ba,x} + F_{Ti,x}+3 F_{O,x} \right ]
\label{Facoustic}
\end{eqnarray}
where $F_{O,x}$ is the average force on the six oxygen ions that are nearest to this Ti ion, since the eigenvector of the acoustic vibration  has an equal weight of $\frac{1}{\sqrt{5}}$ on each ion. A similar computation can be done for the $y$- and $z$-components of these forces, therefore yielding a vectorial force representing the acoustic contribution on each 5-atom unit cell and that is centered on Ti ions.

The results for the $4\times 4 \times 2$ supercell are shown in Figs 1a and 1b for the optical and acoustic contributions, respectively, for a $\theta$ angle of 1$^\circ$, in the bottom BTO layer.  One can see in Figs. 1a and 1b that the force map possesses vortices (indicated in bue and red) at the bottom layer for the optical contribution and that  the acoustic forces also show a vortex topology  with a rotation of the forces being identical to the sense of the rotation of the optical force.  There are also antivortices in Figs. 1a and 1b, in addition to vortices, with the cores of the vortices and antivortices being marked in in green and brown there. It is important to realize that the mean square of the arrows in Fig. 1b is about 3.7 times larger than the one of Fig. 1a, indicating that the acoustic forces are stronger than the optical ones and
hence suggesting that the twisting has more direct consequence on the strains than on the electric dipoles.

{\it Consequences of the acoustic forces on the strain and polarization pattern}

Let us therefore first concentrate on acoustic forces. To better understand and further analyze the effect of twisting on such acoustic forces, we fitted the $x$- and $y$- components of the acoustic forces of Fig. 1b  with harmonic functions and found that

\begin{eqnarray}
F_{acoustic-fit,x} =B_{acoustic,x}+ A_{acoustic,x} cos(2\pi y/L + \phi_{acoustic,x}) \nonumber \\
F_{acoustic-fit,xy} =B_{acoustic,y}+ A_{acoustic,y} cos(2\pi x/L + \phi_{acoustic,y})
\label{harmonic}
\end{eqnarray}
where $A_{acoustic,x}=-A_{acoustic,y}=56.0$ meV/\AA, $B_{acoustic,x}=B_{acoustic,y}=8.1$ meV/\AA, $\phi_{acoustic,x}=-0.80$, $\phi_{acoustic,y}=-0.83$  radians,  and $L$ is the size of the supercell in the $x$- and $y$- directions. We found that the main contribution to these forces comes from $O_{par}$. These resulting  two waves fit rather well the whole pattern of the acoustic forces in the twisted BTO bilayer (see also Ref. \cite{Ghosez}).  Note also that we numerically find (not shown here) that the $A$- parameters increase with the twisting angle. For example, for 3$^\circ$, $A_{acoustic,x}=249$ meV/\AA. One should also note that  a wave compliant with the boundary condition is needed to get Equations (\ref{harmonic}), and the result shows that, among these waves, the one having the maximal possible wavelength (that is equal to the length of the supercell) in the $x$- and $y$- directions, is the one occurring.

Let us now try to relate Eqs.(\ref{harmonic}) to the experimental and theoretical results of Refs. \cite{jorge,flexo}, for the morphology of the shear strains and electric dipoles in the twisted BTO layers. For that, one has  to recall that, in the harmonic approximation,
\begin{eqnarray}
v_x = F_{acoustic-fit,x}/k_{acoustic} \nonumber \\
v_y = F_{acoustic-fit,y}/k_{acoustic}
\label{Hook}
\end{eqnarray}
which is simply the Hooke's law, where $k_{acoustic}$ has a dimension of a spring constant, and {\bf v} is the distortion vector. Note that the evaluation of the spring constant $k_{acoustic}$ was done as follows: in the non-twisted lattice, we slightly moved one specific Ti atom and its Ba and oxygen neighbors by following the acoustic eigenvector. We then project the resulting restoring atomic forces on the acoustic mode, by using again Eq. (\ref{Facoustic}).  The $k_{acoustic}$ spring constant can be obtained from Eq. (\ref{Hook}) by dividing this projected force by the acoustic-related displacement. $k_{acoustic}$ is found to be 4.67 eV/\AA$^2$. Note that we also determined the value of  $k_{acoustic}$ when following the same procedure but starting from the twisted lattice. It was found to be equal to 4.64 eV/\AA$^2$, which is therefore very close to the value of 4.67 eV/\AA$^2$  that we adopt for $k_{acoustic}$.

The displacement ${\bf v}$ described by Eq. (\ref{Hook}) naturally gives rise to the following shear strain component:
\begin{eqnarray}
  \eta_{xy,acoustic-fit}=\frac{d v_x}{d y}+\frac{d v_y}{d x}
  \label{strain}
\end{eqnarray}

Combining Eqs. (\ref{harmonic}), (\ref{Hook}) and (\ref{strain}) yields
\begin{eqnarray}
 & & \eta_{xy,acoustic-fit} =-\frac{2\pi}{kL}[A_{acoustic,x} sin(\frac{2\pi x}{L}+\phi_{acoustic,x})+A_{acoustic,y} sin(\frac{2\pi y}{L}+\phi_{acoustic,y})] \\ \nonumber
 & = &  -\frac{4 \pi A_{acoustic}}{kL}[sin(\frac{\pi (x-y)}{L}+\frac{(\phi_{acoustic,x}-\phi_{acoustic,y})}{2}) cos(\frac{\pi (x+y)}{L}+\frac{(\phi_{acoustic,x}+\phi_{acoustic,y})}{2})]
  \label{shearequation}
\end{eqnarray}

when taking $A_{acoustic}=A_{acoustic,x}=-A_{acoustic,y}$, as for the fit of Fig 1b.

Moreover, flexoelectricity implies that  the gradient of this strain causes the following polarization  in the BTO bilayer \cite{jorge,flexo}:
\begin{eqnarray}
  \delta P_{x,acoustic-fit}=\mu_{xyxy}\frac{\partial \eta_{xy,acoustic-fit}}{\partial y} \\ \nonumber
    \delta P_{y,acoustic-fit}=\mu_{xyxy}\frac{\partial \eta_{xy,acoustic-fit}}{\partial x}
    \label{polarization}
\end{eqnarray}
where $\mu_{xyxy}$ is the shear strain flexoelectric constant of BTO, which is practically equal to 5 nC/m \cite{jorge}.

Plugging Eq.(7) into Eq. (8) gives:
\begin{eqnarray}
 \delta P_{x,acoustic-fit}= -\frac{1}{k_{acoustic}}(\frac{2\pi}{L})^2\mu_{xyxy} A_{acoustic,x}  cos(\frac{2\pi y}{L}+\phi_{acoustic,x}) \nonumber \\
  \delta P_{y,acoustic-fit}= -\frac{1}{k_{acoustic}}(\frac{2\pi}{L})^2\mu_{xyxy} A_{acoustic,y} cos(\frac{2\pi x}{L}+\phi_{acoustic,y})
   \label{polarizationfinal}
\end{eqnarray}

Figures 2a and 2b report the results of Equations (7) and (\ref{polarizationfinal}), respectively, in a (x,y) plane when adopting the $L$, $A$, $\phi$ parameters of Eq. (\ref{harmonic}) fitting Fig. 1b, along with
the $\mu_{xyxy}$ value of BTO reported to be equal to 5 nC/m$^2$ in Refs. \cite{jorge,flexo}.

One can realize that the patterns of Figs 2a and 2b reproduce reasonably well the measurements of Ref. \cite{jorge} as well as the numerical predictions of Ref. \cite{flexo}, which demonstrates that our model starting from the forces for the acoustic displacements, along with the effect of flexoelectricity for the electric dipoles, captures the physics of Moir\'e patterns in twisted oxide layers. In particular, (i) Figure 2a shows that the shear strain adopts four quadrants, with  two (in green) having
vanishing values while the other two (in red and blue) have significant positive and negative large values, respectively;  (ii) Figure  2b displays clockwise and counterclockwise polar vortices along with two types of antivortices; and
(iii) the analytical expressions of Eqs.  (7) and (\ref{polarizationfinal}) are fully consistent with those assumed or determined in Ref. \cite{flexo}.
One can also notice that the value of the maximal strain obtained at the twisting angle of 1$^\circ$ is about 0.7\% in Fig 2a. Based on our aforementioned values of $A_{acoustic}$ for the angles of 1$^\circ$ and 3$^\circ$, we estimate that
such maximal strain should be about 3\% for an angle of  3$^\circ$, which agrees rather well with  the value of 2\% experimentally found  at 3$^\circ$ \cite{jorge}.
Similarly, the local polarization in Fig. 2b is numerically found to have an average magnitude of 4.6 $\mu C/ cm^{2}$ for  1$^\circ$. This polarization should become to be about 20.4 $\mu C/ cm^{2}$ for a twisting angle of 3$^\circ$, which is in remarkable agreement with
the measured data of 20 $\mu C/ cm^{2}$ obtained in Ref.\cite{jorge} for an angle of 3$^\circ$.

{\it Consequences of the optical forces on the polarization pattern}

Let us now  pay attention to the optical forces.  We numerically found that the $x$- and $y$- components  of these optical  forces can be well fitted by:

\begin{eqnarray}
F_{optical-fit,x} =B_{optical,x}+ A_{optical,x} cos(2\pi y/L + \phi_{optical,x}) \nonumber \\
F_{optical-fit,xy} =B_{optical,y}+ A_{optical,y} cos(2\pi x/L + \phi_{optical,y})
\label{harmonicoptical}
\end{eqnarray}

where $A_{optical,x}=-A_{optical,y}=9.1$ meV/\AA,  $B_{optical,x}=B_{optical,y}=-3.7$ meV/\AA, and $\phi_{optical,x}=-0.92$, $\phi_{optical,y}=-0.63$  radians  for the fit of Fig 1a, that is for a twisting angle of 1 $^\circ$.

Using again Hooke's law but now for the motions related to the optical phonons gives:
\begin{eqnarray}
u_x = F_{optical-fit,x}/k_{optic} \nonumber \\
u_y = F_{optical-fit,y}/k_{optic}
\label{Hookoptical}
\end{eqnarray}
where $k_{optic}$ is numerically found to be equal to 3.34 $eV/\AA^2$, following the same procedure as the one mentioned above for $k_{acoustic}$ but now following the eigenvector of the optical soft-mode. The displacements  $u_x$ and $u_y$ are related to the local polarization via:
\begin{eqnarray}
 \delta P_{x,optical-fit}=\frac{Z^*}{a^3_{lat}} u_x  \nonumber \\
  \delta P_{y,optical-fit}= \frac{Z^*}{a^3_{lat}}  u_y
  \label{localmodes}
\end{eqnarray}
with $Z^*$ being the Born effective charge associated with the local modes, which is given to be $9.956e$  for BTO in Ref.  \cite{Vanderbilt-PRB}. Moreover, $a_{lat}$=3.985 \AA~ and is the predicted 5-atom unit cell parameter of BTO bulk in its equilibrium cubic paraelectric phase.

Furthermore, when subtracting  the homogeneous part of the polarization associated with $B_{optical,x}$ and $B_{optical,y}$,  the combination of Eqs. (\ref{harmonicoptical}), (\ref{Hookoptical}) and (\ref{localmodes}) yields:
\begin{eqnarray}
 \delta P_{x,optical-fit}=\frac{Z^*}{k_{optic} a^3_{lat}} [ A_{optical,x} cos(2\pi y/L + \phi_{optical,x}) ] \nonumber \\
  \delta P_{x,optical-fit}= \frac{Z^*}{k_{optic} a^3_{lat}} [ A_{optical,y} cos(2\pi x/L + \phi_{optical,y})]
  \label{polarizationfinaloptical}
\end{eqnarray}

Figure 2c reports such inhomogeneous dipolar pattern calculated by this formula. Interestingly,  (i) the vortices in Fig. 2c are rotating in the opposite sense than in Fig. 2b, and (ii) the average magnitude of the local polarization in Fig. 2c  is about 0.61 $\mu C/cm^2$ that is about 7.5 times smaller than the one in Fig. 2b. 

Let us also provide a word of caution. The much larger predominance of the acoustic forces over the optical ones does not automatically translate into energetic considerations. As a matter of fact, and as indicated by Figs 1c and 1d,
the energy minimum numerically found when moving all the Ti ions in the bottom layer following the optical forces of Fig. 1a is numerically found to be -1.34 meV/u.c., which is only -0.32  meV/u.c.  higher than the energy minimum resulting from the motions of these Ti ions but by following the acoustic forces of Fig. 1b. However and despite this small difference in energy, the energetic hierarchy of these two minima confirms that the system prefers to follow the acoustic forces to develop a striking morphology for the shear strain and then to adopt an array of vortices and antivortices for the electric dipoles thanks to flexoelectricity rather than to follow the optical forces and then generate the original mapping of shear strain via coupling between strains and dipoles. It also worthwhile to indicate that the optical displacement associated with the minimum in energy of Fig. 1c, which is equal to 0.00454 $\AA$, corresponds to  an average magnitude of 1.15  $\mu C/cm^2$ for the local polarization. This value is  about four times smaller than the one of Fig. 2b and one order of magnitude smaller than the one reported in Ref. \cite{jorge}, hinting again that the polar vortices and antivortices seen in BTO twisted layers likely come from a combination of two effects, namely the response of shear strains to twisting then followed by flexoelectricity.

In summary, our first-principles calculations along with an original analysis demonstrate that the  polar Moir\'e patterns in the ferroelectric oxide layers mostly originate from the activation of the forces acting on the {\it acoustic} displacements when these layers are twisted.
Such ``acoustic'' forces naturally result in the appearance of a striking shear strain pattern that can analytically be described by Eq. (7) and which then generate the self-assembled arrays of polar vortices and antivortices thanks to the flexoelectricity expressed by Eqs. (8) and  (9). There is  also  a force acting at the {\it optical} motions that could contribute to the creation of these polar topological defects but at a smaller quantitative extent and in the opposite qualitative fashion.  We hope that the present study deepens the general knowledge of topology and ferroelectrics.

The authors thank the Vannevar Bush Faculty Fellowship (VBFF) Grant No. N00014-20-1-2834 from the Department of Defense and an Impact Grant 3.0 from ARA.
This work was conducted fully or in part with the support of the Arkansas High Performance Computing Center which is funded through multiple National Science Foundation grants and the Arkansas Economic Development Commission. C. P. acknowledges partial support from the U.S. Air Force Office of Scientific Research under grant agreement FA9550-24-1-0263.

\end{document}